\documentstyle[preprint,aps]{revtex}
\begin{document}
\tighten
\draft
\title{Superconducting to normal state phase boundary \\ in arrays of
ultrasmall Josephson junctions\footnote{ To appear in
{\em Proceedings of the NATO
Advanced Research Workshop on Mesoscopic Superconductivity},
F. Hekking and G. Sch$\ddot o$n, eds. Physica B, North Holland (1994).
}}

\author{Jorge V. Jos\'{e}$^{1,2}$ and Cristian Rojas$^{2}$\\
{\it $^1$Instituut voor Theoretische Fysica,\\ Princetonplein 5,
Postbus 80006, 3508 TA Utrecht, The Netherlands\\
$^2$Department of Physics, Northeastern University,\\ Boston
Massachusetts 02115, USA}}
\date{\today}
\maketitle
\begin{abstract}
We study the competition between Josephson and charging
energies in two-dimensional arrays of ultrasmall Josephson junctions,
when the mutual capacitance is dominant over the self-capacitance. Our
calculations involve a combination of an analytic WKB renormalization group
approach plus nonperturbative Quantum Monte Carlo simulations. We consider
the zero frustration case in detail and we are able to make a successful
comparison between our results and those obtained experimentally.

\end{abstract}
\narrowtext

\section{\bf Introduction}
\label{sec:intro}

Recent advances in submicrometer technology have made it possible to
fabricate relatively large arrays of ultrasmall SIS
(superconductor to insulator to superconductor) Josephson
junctions\cite{moij,gee,herre,herre2,tighe,chalmers}. These arrays
consist of Josephson junctions made of superconducting islands separated by
an insulating barrier\cite{moij}. The areas of these junctions can vary
from a few microns to submicron sizes. In the latter case the effective
capacitance of the junctions can be smaller than  femtoFarads (fF=$10\times
10^{-15}$ Farads).
Under these circumstances the long range phase coherent properties of the
Josephson junction arrays (JJA)
depend crucially on the interplay between the Josephson ($E_J$) and
charging energies ($E_C$). In this paper we will be concerned with
studying this interplay for the specific parameter ranges of the experiments
carried out in Delft\cite{herre2}. The charging energy associated with adding
a single charge to an island is $E_{C_{\rm s}}=\frac{e^2}{2C_{\rm s}}$,
whereas the corresponding energy necessary to transfer a charge from an
island to a nearby one is $E_{C_{\rm m}}=\frac{e^2}{2C_{\rm m}}$. Here $e$
is the electronic charge. For the Delft samples, the
self capacitances are  typically $C_{s}\sim 3\times 10^{-18}$F,
while the mutual capacitances are on the order
of $C_{m}\sim 1\times 10^{-15}$F. This means that $C_m$ is
three orders of magnitude larger than $C_s$. Prior to the advent of these
types of JJA, most theoretical studies assumed
that the dominant term was the self-capacitive contribution, since the
calculations are easier and also no significant changes in the
results were expected from having the extra $C_m$ contribution. In the
Delft experiments
a phase diagram was obtained as a function of temperature vs the quantum
parameter
\begin{equation}
\label{alfa}{\alpha\equiv \frac{E_C}{E_J}}.
\end{equation}
The phase diagram that includes the experimental results plus our theoretical
results  is shown in Fig. 1. When $\alpha=0$, the array is modeled by the
two-dimensional XY model and it has been found to be a faithful
representation \cite{bkt2} of the
Berezinskii-Kosterlitz-Thouless (BKT) scenario\cite{bkt,jkkn}. As $\alpha$
increases (in the Delft experiments the values of $\alpha$ cover the
range $0.13$ to $4.55$, while in the Harvard samples the $\alpha$ can be as
large as 33),
the critical temperature decreases monotonically. As $\alpha$ increases
further, at fixed temperature, one is expected to change from a
superconducting phase characterized by quasi-long range phase coherence (SC)
to a charge dominated insulating phase (I). For small and fixed $\alpha$,
as temperature increases, we move from a SC phase to a resistive or
conducting one. For larger values of $\alpha$ the transition is expected to
be between an insulating and a conducting phase.
In the insulator normal phase boundary,
the idea is that in the case when the $C_s=0$, a mutual capacitance model
would map to a two-dimensional Coulomb gas model and thus a charge unbinding
BKT transition would ensue\cite{yoshi,moij2,fazio}. This scenario has not
been found experimentally since in practice, although $C_s$ can be
three orders of magnitude smaller than $C_m$, the electrostatic screening
length is much
smaller than the lattice size and the possible BKT transition is masked.
This means that experimentally the boundary I to N is strictly speaking not
a thermodynamic  phase boundary but a cross over region\cite{tighe,herre2}.
The situation is different in the SC to N boundary
since that is a true thermodynamic phase boundary. An important question
is, however, if the models that have been proposed to describe the JJA
are faithful representations of the experimental systems  studied.
This question is non
trivial since there is evidence for having extraneous elements in the
arrays, e.g. stray charges, that have not been accounted for properly in the
models so far considered. It is for this reason
that we have thought it important to carry out detailed {\bf quantitative}
calculations of the SC to N phase boundary  and to make detailed
comparisons with experiments.
We can already note that our results shown in Fig.1, coming from a
WKB renormalization group calculation,
fit the experimental results quite well in the small $\alpha$ range. On the
other hand, a rather  good fit to the experimental results
comes from carrying out a nonperturbative quantum Monte Carlo  calculation.
In this paper we describe mainly the WKB-RG calculation, and we will briefly
mention the QMC results, which will be discussed in more detail
elsewhere\cite{jens,cristian}.

An important result emerging from the WKB-RG calculation is the possible
existence of a low temperature instability. This $QUIT$ (QUantum fluctuation
Induced Transition) was originally found in a self-capacitive
model\cite{jos84,jacobs}. Here we show that the results are also true in the
case when $C_m$ dominates. This extension is non trivial since the
self-capacitive model {\bf has no} I to N phase boundary with a BKT type charge
unbinding transition, even theoretically. One can then wonder if the
competition between the insulating and
superconducting phases at sufficiently low temperatures may quench the QUIT.
Within the context of the  WKB-RG calculation this is, in fact,
not the case as we discuss in this paper.
{}From the WKB-RG analysis the QUIT instability may be conjectured to be
a phase boundary from a SC to a normal phase\cite{jos84}.

There was early support for the existence of the QUIT in the $C_m=0$,
$C_s\neq 0$ case from nonperturbative quantum Monte Carlo
simulations\cite{jacobs}.
In the $f=0$ case it was found that the helicity modulus had a discontinuity
between two states with {\bf finite} superfluid densities each. This lead to
the surmise that the QUIT is in fact a transition between two superfluid
states, one dominated by thermal and the other by quantum fluctuations.
Later studies considered the case of $f=1/2$, the fully
frustrated limit, which gave a larger discontinuity in the superfluid density
and at higher $T_{QUIT}(f=1/2)$ temperature\cite{jacobs}. As we shall see
from the  WKB-RG calculation described
below, the $T_{QUIT}(f=0)\sim T_{QUIT}(f=1/2)$. However, what was found
in the QMC calculations was that $T_{QUIT}(f=1/2)\sim 10\times T_{QUIT}(f=0)$.
In a recent calculation which uses an improved QMC algorithm, that
reduces the slow approach to equilibrium\cite{novo}, the large
hysteresis loop seen before in the $f=1/2$ case
was suggested to be directly related
to the long lived metastable states seen in the
classical fully frustrated XY model. The authors, however, did not carry out
the calculations to sufficiently low temperature to test the $f=0$, or
$f=1/2$ QUIT. The question then remains as to the physical nature of the QUIT,
if it does exist. From the WKB-RG analysis there is clearly a low
temperature QUIT instability. The known QMC results for the $f=0$,
 $C_m=0$ case suggest a QUIT between two different superconducting states.
Further studies are needed to ascertain the nature of the putative QUIT.
As we also mention later in this paper there is an experimental low
temperature instability that could very well be a manifestation of the QUIT.

Here we shall report on our main WKB-RG and QMC results and sketch how they
were obtained while a fuller more detailed description will appear
elsewhere\cite{jens,cristian}.

The outline of the rest of this paper is the following:
In Sec. II we define  the model studied here and discuss
the path integral representation of the partition function used in our
analysis. In Section III we outline our WKB-RG analysis in the case where the
Josephson energy dominates. We obtain generalized RG equations from which
we obtain the phase diagram, which is then compared with the experimental
results\cite{herre2}, as well as with the Monte Carlo simulation
results\cite{jens}. As we mentioned above, if
$\xi _E$ is large in the insulating phase the charges interact with a
logarithmic potential and a BKT type transition must
follow\cite{yoshi,moij2,fazio}. In section IV we address the
question as to what happens when we perturb the Coulomb gas model ($C_s=0$)
with the Josephson term in the hamiltonian; Is there a QUIT in the charge
dominated phase? We have carried out a WKB-RG perturbative analysis
in the insulating region and found that {\bf there is no analog}
to the QUIT transition found in the SC phase,
at least to the lowest order in the perturbative analysis.

\section{\bf {The model}}

A Josephson junction array can be modelled by a periodic lattice of
superconducting islands separated by insulating barriers. Each island is
characterized by a Ginzburg-Landau order parameter $\Psi (\vec r_i)=
|\Psi _0(\vec r_i)|e^{i \phi (\vec r_i)}$, where $\vec r_i$ is a
two-dimensional vector denoting the position of each island.
Each one of the islands becomes independently superconducting about the
bulk transition temperature $T_{c0}$. When the temperature is lowered further,
the magnitude of the order parameter, $|\Psi _0(\vec r_i)|$, is nonfluctuating
and the onset of long range phase coherence is responsible for the
zero resistance in the arrays. The onset of phase coherence is due to the
tunneling of Josephson currents between the islands and it is characterized
by a sharp drop to the zero resistance state. The onset temperature
in the arrays can be significantly  modified by making the junctions
ultrasmall. When the junction's capacitance is small the charging energy,
i.e. the energy necessary to transfer a Cooper pair between the islands,
can be large to the point where no Cooper pairs can tunnel any more,
and thus the Josephson current can be reduced to the point
where it is completely quenched. The competition between the Josephson
tunneling and the charging energy, in zero field, can be modeled by
the hamiltonian,
\begin{eqnarray}
\label{ha}{\hat {\cal H}}=H_C+H_J
&=&
{{q^2}\over{2}}\sum_{\vec r_1,\vec r_2}\hat n(\vec r_1){\bf {\bf C}^{-1}
(\vec r_1,\vec r_2)}\hat n(\vec r_2)+
E_J \sum_{<\vec r_1,\vec r_2>}[1-
\cos(\hat \phi(\vec r_1)-\hat \phi(\vec r_2))]
\end{eqnarray}
where $q=2e$.  Here $\hat \phi (\vec r_i)$ is the quantum phase operator
while $\hat n (\vec r_i)$
is its canonically conjugate number operator, which measures the excess number
of Cooper pairs in the island placed at $\vec r_i$. These canonically
conjugate operators satisfy the commutation relations
\begin{equation}
[\hat n(\vec r_1),\hat\phi(\vec r_2)]=-i\delta_{\vec r_1,\vec r_2},
\end{equation}
which imply that in the phase representation we can write $\hat n(\vec r_i)=
-i\frac{\partial}{\partial {\phi(\vec r_i)}}$. A Josephson junction is
essentially a capacitor with an insulating barrier. The geometric capacitance
matrix ${\bf C (\vec r_1,\vec r_2)}$ is assumed to include only the island
capacitance with respect to the ground plane, $C_{\rm s}$, and the
mutual capacitance between nearest neighbor islands, $C_{\rm m}$.
In the square array case, of interest here,
we can write the capacitance matrix as
\begin{equation}
\label{cap}{\bf
C(\vec r_1,\vec r_2)} = (C_{\rm s}+4C_{\rm m})\delta_{\vec
r_1,\vec r_2} + C_{\rm m} \sum_{\vec d} \delta_{\vec r_1,\vec r_2+\vec
d}\,\, ,
\end{equation}
where the sum in the second term is over the nearest neighbor
vectors $\vec d$. The electrostatic potential associated with this capacitance
matrix model leads to the screening length
$\xi_E=\sqrt{\frac{C_{\rm m}}{C_{\rm s}}}$.
This means that when $C_{\rm m}=0$ the charges are maximally screened.
In the arrays fabricated so far the screening length, as calculated from
Eq.~(\ref{cap}) is between $15$ and $20$ plaquettes\cite{herre2,tighe}.
This means that screening
effects in these arrays are important and that including $C_{\rm m}$
in the analysis is essential. As mentioned in the introduction, however,
prior to the fabrication of the arrays with
ultrasmall capacitances, many theoretical studies concentrated either on
mean field approaches\cite{simanek,fishman}, which neglect the special
nature of the BKT ordering in two-dimensions, or in 2-D studies with
$C_{\rm m}=0$ \cite{jos84,jacobs}. Initial MC studies of
the off diagonal problem have also appeared\cite{jens2}. Since the
samples fabricated have a typical ratio $C_{\rm m}\sim
 {C_{\rm s}}10^{3}$, it is clear that if one wants to make a direct
 quantitative comparison to
experimental results we need to extend the analysis to include
$C_{\rm m}$ explicitly. This is the purpose of this paper.

We should stress that the model defined by Eqs.(2-4) is not the most general
model one could consider, for it assumes that all the electric fields are
confined to the two-dimensional plane of the array. This may not be quite
the case for real samples but one would expect that in the regime where
$\xi _E$ is sufficiently large this model should be appropriate.
As we shall discuss below the fact that one is able to make successful
comparisons between experiment and theory in the {\it superconducting
to normal} regime leads us to believe that in that regime at least
the model gives a correct representation of the experimental system.

There are other possibly important elements missing in the model studied
in this paper. To wit: (i) We do not consider here the influence of a constant
external magnetic field. The physics in that case, even in the classical
limit, is highly nontrivial and requires a special treatment\cite{ramirez}.
(ii) We do not include the self-induced magnetic fields, that can also lead to
interesting new physics\cite{daniel}. The reason is that  the
critical currents in the
arrays are found to be too low to lead to any significant magnetic field
screening effects. (iii) We do not include quasi-particle
dissipation. Here again it has been
found experimentally that in the SC to N phase these effects are not important
\cite{moij2}. A low temperature study that includes
quasi-particle dissipation for the diagonal capacitance matrix model
has shown that dissipation counteracts the charging effects so as to
strengthen the Josephson tunneling\cite{choi}. (iv) We could also
consider the effect of disorder, in particular the one due to random stray
charges in the array. These random frustration charges may also be more
important in the insulating phase than in the superconducting one, which is
the one we concentrate on in this paper.


Here we are interested in calculating the thermodynamic properties of the
model defined by $\hat {\cal {H}}$. The corresponding partition function is
defined by
\begin{equation}
Z = {\rm Tr}\left\{ e^{-\beta\hat {\cal H}}\right\}\label{zeta}.
\end{equation}
The trace is taken over  $\hat \phi$ or $\hat n$. It is convenient, for
calculational purposes, to change from the operator expression to the
imaginary-time Feynmann path integral representation of $Z$. We accomplish
this by using
\begin{equation}
           <n(\vec r_1)|\phi(\vec r_2)> = \delta_{\vec r_1,\vec r_2}
           \frac{\exp\{ i n(\vec r_1)\phi(\vec r_1)\}}{\sqrt{2\pi}}.
\end{equation}
For this problem we follow steps parallel as those described in Ref.
{}~\cite{jacobs}.
This means that we discretize the imaginary-time axis into
$L_\tau$ time slices separated by a distance $\epsilon =
\frac{\beta \hbar}{L_\tau}$. We  recover, in principle, the fully
quantum results in the limits $L_\tau\to\infty$, $\epsilon\to 0$ with
$\beta \hbar$ kept fixed. Following Ref.~\cite{jacobs} we get  the
partition  function expression (up to order $O(1/L_{\tau})$)
\begin{equation}
Z = \prod_{\tau=0}^{L_{\tau}-1} \sqrt{{\rm det}[ C ] }
    \prod_{\vec r} \int_{0}^{2\pi}\!\!\sqrt{\frac{L_{\tau}}
    {2\pi\beta q^2} } d\phi(\vec r,\tau)\!\!\!\!\sum_{m(\vec r,\tau)
     = -\infty}^{\infty}\!\!\!\!\exp\bigg[-\frac{1}{\hbar}S[\{\phi\},\{m\}]
     \bigg]
     \label{eq(2.3)}
\end{equation}
where we defined the action
\begin{eqnarray}
\frac{1}{\hbar}S[\{\phi\},\{m\}] = \frac{ \beta}{L_{\tau}}
     \sum_{\tau = 0 }^{L_{\tau}-1} \Bigg[& &H_J(\{\phi(\vec r,\tau)\})
     -\frac{L_{\tau}}{2\beta q^2}\!\!\!\sum_{<\vec r_1,\vec r_2>}
       [\phi(\vec r_1,\tau\! +\! 1) - \phi(\vec r_1,\tau)
       + 2\pi m(\vec r_1,\tau) ] \times \nonumber \\
       & &\times{\bf C}(\vec r_1,\vec r_2)[\phi(\vec r_2, \tau\! +\! 1)
       -\phi(\vec r_2,\tau) + 2\pi m(\vec r_2,\tau)]\Bigg] +
               \nonumber \\
               & &+ O(1/L_{\tau}).
               \label{eq(2.4)}
\end{eqnarray}
Here $H_J(\{\phi(\vec r,\tau)\})$ is the Josephson hamiltonian in terms
of the phase variable $\phi(\vec r,\tau)$.
The quantum nature of this c-number functional integral comes from imposing
the periodic boundary condition
\begin{equation}
\phi(\vec r,L_\tau) = \phi(\vec r,0).       \label{eq(2.5)}
\end{equation}
These are the basic equations used in the  WKB-RG and Monte Carlo
calculations described in the next section.


\section{\bf {WKB-RG study about the SC to N phase boundary}}

Our approach here is to perturb the physics described by
$H_J(\{\phi(\vec r,\tau)\})$, in the limit when the charging energy is
small. When $\alpha=0$, $H_J$ describes the physics of the classical
2-D XY model\cite{bkt,jkkn}. In this case we have the BKT scenario that
depends on the
thermal nucleation of vortex-antivortex pairs (VAP). The density of
VAP increases exponentially as the temperature rises until they unbind
at the critical temperature
$T_{\rm BKT}(\alpha=0)=T_{\rm BKT}^{(0)}= (\pi E_J/2k_{\rm B})$.
The BKT scenario is best understood in terms of a renormalization
group (RG) analysis\cite{bkt,jkkn}. The RG
flow diagram is obtained from a perturbation expansion in powers of the
vortex pair density $y=y_0e^{-\frac{\pi ^2}{2}K}$. Here $K=\beta E_J$ and
$y_0$ is the initial condition for the bare vortex pair density.
In the standard BKT picture there is a line of fixed points for $0\leq T\leq
T_{\rm BKT}^{(0)}$, with algebraically decaying correlation functions.
In the self-capacitive model, at $T=0$, one can map the problem to an
anisotropic three-dimensional XY model, which must have a standard phase
transition at a critical value of $\alpha ^c_s$. Around the $T=0$
critical point $\alpha_s^c$, one expects to have exponentially decaying
correlation functions while at $\alpha_s^c$ the correlations must decay
algebraically. The important question
is then: How do we go  from the BKT regime, with algebraic decaying correlation
functions to the very low temperature exponentially decaying one? There must
be a discontinuity in going from one limit ($\alpha _s=0$, $T\neq 0$) to the
other one ($\alpha _s\neq 0$, $T=0$). Below we discuss
the evidence we have found, including a possible experimental candidate,
that there may indeed be a QUIT at low temperatures.

The situation  when $C_s=0$ and $C_m\neq 0$  is actually
quite different. In that case the $T=0$ limit can be approximately
represented by two
coupled three-dimensional XY models, one describing the phase degrees of
freedom and the other the charges. As a function of $\alpha_m$ we can go
from a phase dominated region, with a 3-D type XY model critical properties
to one dominated by a 3-D Coulomb gas.
There is not much known about the critical properties,
and in particular the correlation function behavior, of the two coupled
3D XY models. So, strictly speaking, we can not state what kind of crossover
we should expect when going from $T=T_{BKT}(\alpha _m,\alpha_s)$ to $T=0$.
Some understanding of the physics in this limit can be obtained by using the
Villain transformation, both for the charging energy term and for the phase
contribution\cite{fazio}. The Villain approximation is, however, valid only in
a restricted range of $\alpha$ values which do not cover the full experimental
range. One could conjecture, however, that the properties of the
Villain approximated models is in the same universality class as the full
coupled XY models, which is in fact the case when $\alpha=0$ \cite{jkkn},
but this needs to be explicitly shown.
Furthermore, the general case treated here where both $C_s\neq 0$ and
$C_m\neq 0$ is more complicated since the effective Coulomb gas in the
insulating phase has a finite screening length. All these issues need to be
studied further.

To find  the corrections to the BKT scenario due to the charging effects
we carry out a semiclassical or WKB analysis of the model. This was
originally done for the self-capacitive model in Ref.~\cite{jos84}.
Here we follow a similar approach, except that technically the problem is
more demanding.

To evaluate the partition function in the SC to N, or small $\alpha$ regime, we
notice that we can extend the range of integration of the phases in
Eq.~(\ref{eq(2.3)}) from
[0,$2\pi$] to [$-\infty,\infty$], while at the same time all but one of
the summations over the set of $\{m\}$'s can be eliminated.
The resulting expression for the partition function, in the
$L_\tau\rightarrow\infty$ limit, then reads
\begin{eqnarray}
           Z = \sqrt{{\rm det}[{\bf C}]} \int_{0}^{2\pi}
               \prod_{\vec r}
               \sqrt{\frac{L_{\tau}}{2\pi \beta q^2}} d\phi(\vec r,0)
               \!\!\!\!\sum_{m(\vec r )=-\infty }^{\infty} \int_{-\infty}^
               {\infty} & &\prod_{\tau = 0}^{L_{\tau}-1}\!\!\sqrt{{\rm det}
               {\bf [C]}}
               \prod_{\vec r} \sqrt{\frac{L_{\tau}}{2\pi \beta q^2}}
               d\phi(\vec r,\tau) \times \nonumber \\
               & &\times\exp\bigg[-\frac{1}{\hbar}\int_{0}^{\beta\hbar}
               d\tau L_E \bigg],
           \label{eq(3.1)}
\end{eqnarray}
where the euclidean lagrangian is
\begin{equation}
L_E = \frac{1}{2}\left(\frac{\hbar}{q}\right)^2\sum_{<\vec r_1,\vec r_2>}
\frac{d\phi}{d\tau}(\vec r_1,\tau){\bf C}(\vec r_1,\vec r_2)\frac{d\phi}
{d\tau}(\vec r_2,\tau) + H_J(\{\phi\}).
\label{eq(3.2)}
\end{equation}
The boundary condition, Eq.~(\ref{eq(2.5)}), now reads
\begin{equation}
\phi(\vec r,\beta\hbar) = \phi(\vec r,0) + 2\pi m(\vec r),
\label{eq(3.3)}
\end{equation}
where the $\{m(\vec r)\}$'s are the winding numbers. We note that since
the lagrangian is invariant under the transformation
$\phi(\vec r,0)\rightarrow\phi(\vec r,0) + 2\pi l(\vec r)$
for all integers $\{l(\vec r)\}$, we can extend the limits of
integration over $\phi(\vec r,0)$ to $[-\infty,\infty]$,
the difference coming  only  from an overall multiplicative constant.
Now that the limits of Eq.~(\ref{eq(3.1)}) are all from $[-\infty,\infty]$ we
can make the following change of variable\cite{schulman}
\begin{equation}
\phi(\tau,\vec r) = \frac{2\pi}{\beta\hbar}m(\vec r)\ \tau\ +\
                    \overline{\phi}(\vec r)+\phi_f(\vec r,\tau).
\label{eq(3.4)}
\end{equation}
Here $\phi_f(\vec r,\tau)$ represents the quantum fluctuations of
the path about its mean value $\overline{\phi}(\vec r)$. These
quantum fluctuations become larger than the thermal ones as $\alpha$
increases or as the temperature decreases. This
means that we would need to take higher order harmonics in the Fourier
series into account when the quantum effects are not relatively small.
Because of  the periodicity in Eq~(\ref{eq(3.3)}),
 $\phi_f(\vec r,\tau)$  can be expanded in the Fourier series
\begin{equation}
\phi_f(\tau,\vec r) = (\beta\hbar)^{-1/2} \sum_{k=1}^{\infty}[\phi_k(\vec r)
e^{i\omega_k\tau}+C.C.],
\label{eq(3.5)}
\end{equation}
where the $\omega_k=2\pi k/\beta\hbar$ are the Bose Matsubara
frequencies. Substituting Eqs.~(\ref{eq(3.4)}) and ~(\ref{eq(3.5)}) in
Eq.~(\ref{eq(3.1)}),  expanding the Josephson term up to second order in
$\phi_k(\vec r)$, i.e. up to  order $O(q^2)$ or equivalently to $O(\alpha)$,
we obtain an effective  action for the classical
variables $\overline\phi(\vec r)$ after evaluating the integrals. In obtaining
the effective action we note that once the integrations over the
$\phi_k(\vec r)'s$ are carried out, the partition function still includes
a summation over the $m(\vec r)$'s.
In the semiclassical limit the contributions to the partition function
from configurations with $m(\vec r)$ different from zero are exponentially
small, so that we can safely take $m(\vec r)=0$ for all $\vec r$.
A most important property of the Josephson hamiltonian
$H_J$ is that it is a periodic function of its argument. This implies that
in the expansion the second order derivative with respect to the argument
in $H_J$ is proportional to $H_J$ itself. Specifically, for the cosinusoidal
form of $H_J$ we have $H_J'' = -H_J$ + constant. This important property of
$H_J$ allows us to write the effective partition function as a 2-D classical
XY model with an effective  coupling constant. The effective partition
function to this order of approximation is\footnote{ We must note that in
obtaining the effective partition function we assumed that the phases
took values between $[-\infty,\infty]$, whereas in the classical XY model
the phases are constrained to lie in the $[0,2\pi]$ range. Following this
route makes the derivation of the effective action  more direct. However,
in the small $\alpha$ regime of interest here the differences between
the two ranges for the phases can be shown to be exponentially small.}
\begin{equation}
Z_{eff} = \int \prod_{\vec r} \frac{d\overline\phi(\vec r)}
{2\pi} \exp[-\beta_{\rm eff}H_J(\overline\phi)],
\label{eq(3.6)}
\end{equation}
where the effective inverse temperature is given by
\begin{equation}
\beta_{\rm eff} = \beta -\frac{(q\beta)^2)}{12}\left[{\bf C}^{-1}(0)-
{\bf C}^{-1}(\vec d)\right].
\label{eq(3.7)}
\end{equation}
Note that we have explicitly used the fact that  ${\bf C}(\vec r_1,\vec r_2)
={\bf C}(|\vec r_1 -\vec r_2|)$, valid for a periodic lattice,
so that we can  Fourier transform the capacitance matrix.

Once we have a hamiltonian which is just like the 2-D classical XY model,
we can write down the corresponding effective RG recursion relations
to lowest order in $x$ as
\begin{eqnarray}
\frac{dK}{dl} &=& 4\pi^3 K^2 {\tilde y}^2 \frac{(1-xK)^2}{(2Kx-1)},
\label{eq(3.9)}\\
\frac{d\tilde y}{dl} &=& [2-\pi K(1-xK)]\tilde y.
\label{eq(3.10)}
\end{eqnarray}
In writing these equations we defined the variables
\begin{eqnarray}
x &=& \frac{q^2}{12E_J}\left[{\bf C}^{-1}(0)-{\bf C}^{-1}(\vec d)\right],\\
\tilde y &=& \exp[-{\frac{\pi^2}{2}K(1-xK)}]\equiv
\exp[-{\frac{\pi^2}{2}K_{eff}}].
\label{eq(3.8)}
\end{eqnarray}
The variable $x$ is the $\alpha$ parameter when the capacitance matrix is
not just the self or the mutual capacitance.
The RG equations are solved using as initial conditions
$K_{eff}(l=0)\equiv K_{eff}^0$ and $\tilde y (l=0)\equiv \tilde y_0$.
As written, the RG equations are valid for an arbitrary ratio between the
self and the mutual capacitances.
We first notice that in the $x=0$ limit the RG equations reduce to
the standard Kosterlitz RG equations\cite{bkt,jkkn}, as they should.
The form of the vortex density $\tilde y$ is
most important. As
mentioned above, $\tilde y(x=0)$ grows exponentially with temperature.
When $x\neq 0$ and as a function of temperature, $\tilde y$
exhibits a low temperature minimum. This is shown in the  discontinuous line
in Fig. 2. This behavior for $\tilde y$ is easy to understand physically.
At high temperatures the difference between $\tilde y$ and $y$ is very small.
However, below the minimum the increase in the vortex pair density is
due to nucleation of VAP via quantum phase slips.
Of course, we need to remember that we have done a perturbative
calculation in $x$ and therefore
we may not be on safe ground when $\tilde y$ starts increasing again
at low temperatures. Nevertheless, as often happens with WKB derived results,
the fact that the perturbative analysis shows a low temperature instability
is likely to be true. In fact
we also have found numerical evidence for the low temperature instability in
our QMC calculations\cite{jacobs}.

The RG equations have two nontrivial fixed points, one that
corresponds to the effective BKT thermal fluctuations driven transition,
and the other that corresponds to the
QUantum fluctuations Induced transition (QUIT)\cite{jos84,jacobs}.

The RG level curves in the ($\tilde y, K$) phase space, to lowest
order in $x$, result from solving
Eq.~(\ref{eq(3.8)}) and
\begin{equation}
\pi x K -\pi {\rm ln}K -\frac{2}{K} + 2\pi^3 {\tilde y}^2 = A,
\end{equation}
for different initial conditions. Figure 2 shows the RG flows for
different initial conditions starting with different
values for ($\tilde y_0, K^0$)
along the discontinuous line in the figure. Each RG flow line corresponds
to a different temperature with the arrows indicating the direction of
increasing $l$. We clearly see from the figure that we can divide the
temperature axis into three different regions. In the region between
$[K^{-1}_{QUIT},K^{-1}_{BKT}]$, as the value of $l$ increases we eliminate
VAP, with the unusual property that the vortex density can initially
grow for a while
before tumbling to the critical line $\tilde y=0$. This means that in this
region at $l=\infty$ there are no VAP with infinite separation, i.e.
unbounded. Below $K^{-1}_{QUIT}$, as $l$ increases, the RG trajectories grow
away from the $\tilde y=0$ line, nonmonotonically, indicating that the
perturbation expansion in $\tilde y$ is no longer valid.
If we associate the instability in the perturbation theory
with the normal state behavior, one could say then that
this behavior is characteristic of a reentrant phase transition, i.e.
going from N to SC to N. This is certainly  the case in the
high temperature regime but not necessarily so at low temperatures. The
single line that divides the two types of behaviors mentioned above is the
separatrix that determines the critical temperature. This is the line with
 the highest temperature for which we can touch the $\tilde y=0$
line. The corresponding separatrix value of the constant $A=A_c$ is determined
from the condition that it passes through
the point $(K^{-1}_{BKT},\tilde y=0$). This leads to the result that the
critical point is obtained from solving the equation,
\begin{equation}
 \pi xK_c-\pi{\rm ln}K_c-\frac{2}{K_c}
 2\pi^3 \exp\{-\pi^2 K_c(1-xK_c) \} =A_c(x),
\end{equation}
where $A_c(x)$ is obtained from  $A_c(x)=2\pi x K_c-\pi -\pi
\rm {ln} K_c$, and $K_c$ is the solution to the equation
$2=\pi x K_c(1-xK_c)$. We'll come back to the problem of determining
$K^{-1}_{BKT}$ in the next section, where we make a comparison with the
experimental results. It is known, however,  that the determination of
$T_{BKT}$ using the RG equations is not quantitatively
exact, since the RG analysis is explicitly derived  for the Villain
action. In comparing with experiment in the next section we
will take this into account. Here we  present the corrections to
the classical results to the leading order in $x$, which give the correct
qualitative trends. Specifically, expanding in powers of $x$ we find that
$T_{BKT}$ and $T_{QUIT}$ are given by
\begin{eqnarray}
T_{\rm BKT}&\approx& T_{\rm BKT}^{(0)} - \frac{E_J}{k_{\rm B}}x + O(x^2),
\label{eq(3.11)}\\
T_{\rm QUIT}&\approx& \frac{E_J}{k_{\rm B}}x+O(x^2).
\label{eq(3.12)}
\end{eqnarray}

Notice that these equations are applicable not only in 2-D,
for if the system described by Eq.~(\ref{eq(3.6)}) has a
transition point at some $K_{\rm eff}^{\rm c}$ then the equation
$K_{\rm eff}^{\rm c}~=~K-xK^2$ has two solutions for $K$, which are the ones
implied in Eq.~(\ref{eq(3.10)}). Moreover, notice that the results to the
first order in $x$ are independent of the specific value
of $T_{\rm BKT}^{(0)}$.
This means that if we consider the finite magnetic field case,
the corresponding critical temperature will be $T_c(B)\approx
T_c^{(0)}(B)-(E_J/k_{\rm B})x+O(x^2)$. Furthermore, we notice that,
to the lowest order in $x$,
the $T_{QUIT}$ must be the {\bf same} with and without a field.
This fact will be compared with the  experiments in the next section.

The explicit leading order calculation of the correction to the BKT critical
temperature in the asymptotic limits in which either the self or the mutual
capacitance dominates results in
\begin{equation}
\frac{T_{\rm BKT}}{T_{\rm BKT}^{(0)}} = \left\{
\begin{array}{ll}
    1-\frac{4}{3\pi}\frac{E_{C_{\rm s}}}
    {E_J}+O\left[\left(\frac{E_{C_{\rm s}}}
    {E_J}\right)^2\right],
    & \mbox{if $C_{\rm s}\gg
     C_{\rm m}$}\\
    1-\frac{4}{3z\pi}\frac{E_{C_{\rm m}}}
    {E_J}+O\left[\left(\frac{E_{C_{\rm m}
    }}{E_J}\right)^2\right],
    & \mbox{if $C_{\rm s}\ll
    C_{\rm m}$}
     \end{array}
     \right.
     \label{eq(3.14)}
\end{equation}
here $z$ is the coordination number of the array, and for a square array in
two dimensions, $z=4$.

\section{\bf {Comparison to experiment}}

We now move on to a brief discussion of how  we obtained the results presented
in Fig. 1. As mentioned before, in trying to find quantitative correspondence
between experiment and theory it is important to ascertain the validity
of the theoretical models employed to study the arrays. We have carried
out two different types of checks. One based on the RG analysis described
in the previous sections and the other from a nonperturbative quantum
Monte Carlo calculation\cite{jens,cristian}.
We discuss  the RG analysis here and only
briefly mention the QMC results, with more details left for
forthcoming publications \cite{jens,cristian}.

As mentioned  in section III , it is known that the RG equations
do not lead to quantitatively exact results for the critical temperatures.
They do, of course, lead to the correct universal
critical  exponents. However, what has so far been measured experimentally
is the phase boundary between the SC and the N phases. We then need  a
consistent way to compare the RG results with the experimental results.

We first note that the phase diagram of Fig. 1 is plotted as a function of
$\alpha_m=\frac{E_{C_m}}{E_J}$, since experimentally $\alpha_s$ is
three orders of magnitude smaller. We can then write $K_{eff}=K(1-xK)$
with $C_S=0$ as
\begin{equation}
K_{eff} = K - \frac{\alpha_m}{6}K^2.
\end{equation}
Next we  set the critical temperature for the classical model to be the one
obtained in classical MC simulations (e.g. \cite{ramirez})
$1/K_c^{(0)}\approx 0.93$, so that the critical temperature $T_c(\alpha_m$)
of the actual model is given by the equation
\begin{equation}
k_BT_c(\alpha_m) = \frac{1}{2K_c^{(0)}} \left[1+\sqrt{1
-\frac{2\alpha_m}{3}K_c^{(0)}}\right].
\end{equation}
To lowest order  in $\alpha_m$ this equation gives
\begin{equation}
T_c(\alpha) = T_c(0) - \frac{\alpha_m}{6},
\end{equation}
whereas the maximum value of $\alpha_m$ for which there is a
physical solution is $\alpha_m = \frac{3}{2} k_BT_c(0)
\approx 1.4$. The results
obtained from this analysis are shown as a discontinuous line in Fig. 1. By
following this approach we see that for $\alpha_m\leq 1$ the RG result
is actually quite good when compared to the experimental and the
MC results.

We also have  extended our previous QMC calculations to the case when
the off-diagonal capacitance is dominant. The results are shown in Fig. 1
by the crosses, including their error bars. It is clear from these results
that the correspondence between experiment and QMC results is excellent,
up to nonperturbative values of $\alpha_m$. This leads us to the conclusion
that the model studied here does provide a good representation of the
experimental system, at
least in the SC to N regime. In the next section we will briefly discuss
what happens in the insulating region to normal region.

The discussion presented above dealt with the SC to N phase boundary.
What about evidence for a QUIT? In this regard we note that in the
experimental results of Ref.~\cite{herre2} there are results
for a sample with a nominal
$\alpha_m=1.67$ for which there is a double type of reentrant behavior.
The low temperature glitch seen in the resistance versus temperature diagram
occurs at $T'=40mK$. Moreover the latter instability is also characterized by
an increase in noise fluctuations in the IV characteristics measured
in this sample. If we assume that what is seen at $T'$ is related to the QUIT,
using Eq.(24) and the parameter values of the experiment we get a
$T_{QUIT}=33mK$, rather close to the experimental value. Furthermore
when the same experiment is repeated in a small magnetic field the low
temperature instability is found at he same temperature, i.e.;
$T'(f=0)=T'(f=0.08)$. This result is also consistent with
Eq. (24) which also leads
to a $T_{QUIT}$ independent of $f$ at leading order in $\alpha_m$.
These results may just be coincidental and more work needs to be done
to conclusively connect the $T_{QUIT}$ with the low temperature
instability already seen in the Delft experiments.

\section{\bf {Insulating to Normal cross over}}

As mentioned in the introduction, as $\alpha_m$ increases there is a SC to I
transition at finite temperature. In the $\alpha_s=0$ case,
the insulating phase has been modelled as a two-dimensional Coulomb
gas of charges with a possible BKT charge unbinding
transition\cite{yoshi,moij2,fazio}. This
situation has been studied extensively, in particular in Ref.~\cite{fazio}.
Here we ask if there is an equivalent QUIT in the
insulating phase at low temperatures. If we draw an analogy to
the quantum induced vortices in the SC phase one could
also imagine that the the number of free dipoles in the arrays could increase
due to quantum fluctuations. However, as we show below the vortices and
charges are not dual to each other in that sense.

The calculation described here aims at finding the leading correction
to the charging hamiltonian due to Josephson junction fluctuations.
 We then expand the Josephson contribution to $Z$ as
 \begin{equation}
\exp\left[-\frac{1}{\hbar}\int_{0}^{\beta\hbar}d\tau H_J(\tau)\right]\approx
1-\frac{1}{\hbar}\int_{0}^{\beta\hbar}d\tau H_J(\tau)+\frac{1}{2\hbar^2}
\int_{0}^{\beta\hbar}\int_{0}^{\beta\hbar}d\tau d\tau' H_J(\tau)H_J(\tau')
+\dots
\label{eq(4.1)}
\end{equation}
As before, we use  Eqs.~ (12) and (13), but
this time we integrate out both $\phi_f(\tau,\vec r)$
and $\overline{\phi(\vec r)}$, which leaves us with an
effective action for the $\{m\}$'s,
\begin{equation}
Z\approx Z_\phi \prod_{\vec r} \sum_{m(\vec r)=-\infty}^{\infty}
\exp\left[-\frac{1}{4\tilde K_{\rm eff}}\sum_{\vec r,\vec d}\left(m
(\vec r+\vec d) - m(\vec r)\right)^2\right],
\label{eq(4.2)}
\end{equation}
where we have assumed that $C_{\rm s}\ll C_{\rm m}$. The function
$Z_\phi$ does not contain the $\{m\}$ variables, and the effective
coupling constant $\tilde K_{\rm eff}$ is
\begin{equation}
\tilde K^{-1}_{\rm eff} = \tilde K^{-1}\left[1+\left(\frac{2\pi^2}{\alpha_m}
\right)^2 g(\tilde K)\right],
\label{eq(4.3)}
\end{equation}
given as a function of $\tilde K=(\beta E_{C_m})/2\pi$.
As in the vortex dominated case, we have ended up with a Coulomb gas problem
but with a renormalized coupling constant. The
function $g(\tilde K)$ determines the importance of the zero point
fluctuations of the phases on the charge dominated phase.
The function $g(\tilde K)$  is defined as
\begin{equation}
g(\tilde K)=\int_{0}^{1/2} dt[1-\cos(2\pi t)]\exp\left[-\frac{(2\pi)^2}{z}
\tilde Kt(1-t)\right].
\label{eq(4.4)}
\end{equation}
For a BKT type phase transition we have
 $\tilde  {K}_{eff}=K_{\rm BKT}^{c} = 2/\pi$ \cite{fazio}. It is important
 here  to see if, within this approximation, the system shows a
QUIT or a reentrant transition. We then need  to study the number of
solutions to this equation. From the fact that $K_{\rm BKT}^{c}\sim O(1)$
we can see that for any value of $1/{\alpha_m}$ the function $g(\tilde K)$
has  only one solution for $\tilde K_c$. This means that, to this order
of approximation, {\bf there is no}
QUIT in the charge dominated phase in this model.

A question that immediately arises is: Why is there a difference between the
vortex and  charge dominated phases, in particular in view of the duality
between the two phases extensively studied in \cite{fazio}? The reason is that
the duality is not exact since there is a term in the action, obtained using
the Villain approximation, that breaks this symmetry. If one includes this
term we then see that the cost of producing quantum fluctuations in the
vortices is bounded from above whereas the corresponding
cost in the charge dominated phase is unbounded. In our calculation we
have kept these contributions  intact.

\acknowledgments
We thank  H. van der Zant for many useful discussions and for providing us
with copies of his unpublished work prior to publication.
We also thank J. Houlrik and M. Novotny
for useful conversations and suggestions in the early stages of this work.
This work has been partially supported  by $NSF$ grant DMR-9211339.
%

%
%

\newpage

\begin{figure}
\caption{ Temperature vs. charging energy phase diagram.
The experimental results are denoted by squares. The RG results are
given by the discontinuous line while the quantum Monte Carlo results
are given by the crosses and joined by a continuous line as a guide
to the eye. The latter results include the statistical error
bars in the calculations. }
\label{fig1}
\end{figure}

\begin{figure}
\caption{Renormalization group flow diagram. The discontinuous line indicates
the vortex pair density as a function of temperature. See text for a
discussion  of the analysis of this diagram.}
\label{fig2}
\end{figure}
\end{document}